\documentclass[12pt]{article}
\usepackage{epsfig}

\newcommand{\bea}{\begin{eqnarray}}
\newcommand{\eea}{\end{eqnarray}}
\def\e{{\rm e}}

\newcommand{\gm}{\gamma}
\newcommand{\Gm}{\Gamma}
\newcommand{\dl}{\delta}

\newcommand{\ep}{\epsilon}

\newcommand{\dd}{\mbox{d}}

\newcommand{\ux}{\underline{x}}
\newcommand{\nn}{\nonumber}

\newcommand{\un}{\underline{n}}

\newcommand{\cdo}{\!\cdot\!}
\newcommand{\lra}{\leftrightarrow}

\begin{document}

%%%%%%%%%%%%%%%%%%%%%%%%%%%%%%%%%%%%%%%%%%%%%%%%%%%%%%%%%%%%

\title{\vskip-3cm{\baselineskip14pt
\centerline{\normalsize \hfill DESY 03-081}
\centerline{\normalsize \hfill July 2003}
}
\vskip1.5cm
Solving Recurrence Relations for Multi-Loop Feynman Integrals }
\author{Vladimir A. Smirnov$^a$ and Matthias Steinhauser$^b$\\
{\small $^a$
Skobeltsyn Institute of Nuclear Physics,
Moscow State University,}\\
{\small 119992 Moscow, Russia}\\
{\small $^b$ II. Institut f\"ur Theoretische Physik, Universit\"at
Hamburg,}\\
{\small Luruper Chaussee 149, 22761 Hamburg, Germany}
}

\date{}

\maketitle

\thispagestyle{empty}

\begin{abstract}
We study the problem of solving integration-by-parts recurrence relations for
a given class of Feynman integrals which is characterized by
an arbitrary polynomial in the numerator and arbitrary integer powers of
propagators, {\it i.e.}, the problem of expressing any Feynman integral from
this class as a linear combination of master integrals. We show how the
parametric representation invented by Baikov~\cite{Bai1} can be used to
characterize the master integrals and to construct an algorithm for evaluating
the corresponding coefficient functions. To illustrate this procedure we use
simple one-loop examples as well as the class of diagrams appearing in the
calculation of the two-loop heavy quark potential.

\medskip

\end{abstract}

\newpage

%%%%%%%%%%%%%%%%%%%%%%%%%%%%%%%%%%%%%%%%%%%%%%%%%%%%%%%%%%%%

\section{Introduction}

In the recent years the art of evaluating multi-loop Feynman
integrals has been driven to an impressive high level and very
sophisticated methods and tools have been developed in order to cope with
the
enormous complexity one encounters at higher orders
(see, e.g., the recent reviews~\cite{reviews}).
A standard practical problem is the evaluation of a
class of Feynman integrals corresponding to a given graph.
The integrals differ by the integer powers of the
propagators and the polynomials in
numerators. Most methods heavily rely on the use of recurrence
relations derived with the help of integration-by-parts
(IBP)~\cite{IBP} identities. The recurrence relations are used to
express a complicated integral in terms of simpler ones and, after
repeated use, finally in terms of a small set of integrals,
so-called master integrals, which can not further be reduced. Only
for the latter a complicated integration is possibly necessary.

Any dimensionally regularized Feynman integral corresponding to a given
graph,
with the space-time dimension $d=4-2\ep$ as a regularization parameter
\cite{dimreg},
can be represented in the form
\begin{eqnarray}
  F(\underline{n}) &=&
  \int \cdots \int \frac{\dd^d k_1\ldots \dd^d k_h}
  {D_1^{n_1}\ldots D_N^{n_N}}\,,
  \label{eqbn}
\end{eqnarray}
where $k_i$, $i=1,\ldots,h$, are loop momenta, $n_i$
are integer indices,
underlined letters denote multi-indices, {\it i.e.},
$\underline{n}=(n_1,\ldots,n_N)$,
and the denominators are given by
\begin{eqnarray}
  D_a&=&\sum_{i\geq j \geq 1} A^{i j}_a \, p_i \cdot p_j - m_a^2 \,,
  \label{denom}
\end{eqnarray}
with  $a=1,\ldots,N$.
Irreducible numerators which cannot be linearly expressed through
the given set of the denominators $D_a$ of the propagators are
naturally treated as extra
denominators raised to a negative power. The momenta $p_i$ are either
the loop momenta $p_i=k_i, \; i=1,\ldots,h$, or external momenta
$p_{h+1},\ldots$ of the graph
so that the denominators can be either quadratic or
linear\footnote{Linear denominators usually appear in asymptotic
  expansions of Minkowski space Feynman integrals
  within the strategy of expansion by regions \cite{BS}.}
in the loop momenta $k_i$.

The IBP identities are obtained by applying the operator
$ (\partial/\partial p_i)\cdot p_j$ ($i=1,\ldots,h$ and
$j=1,\ldots,h,\ldots$) to the
integrand in Eq.~(\ref{eqbn}). Due to the properties of dimensional
regularization the resulting integral is zero which
can symbolically be written as
\begin{eqnarray}
  P_{ij}(I^+,I^-)F(\underline{n}) &=&0 \,.
  \label{eqrr}
\end{eqnarray}
$P_{ij}$ are polynomials in
the operators that increase and lower indices, ${\bf I}^+_a$ and
${\bf I}^-_a$, where
\begin{eqnarray}
  {\bf I}^+_a F(\ldots, n_a,\ldots ) &=& F(\ldots, n_a+1,\ldots)
  \,,
  \nonumber \\
  {\bf I}^-_a F(\ldots, n_a,\ldots)&=& F(\ldots,n_a-1,\ldots)
  \,.
\end{eqnarray}

The basic idea is to use the relations of Eq.~(\ref{eqrr})
in order to express
any integral with a given value of the multi-index
$\underline{n}$ in terms of simpler
integrals corresponding to some finite family
of multi-indices $\underline{n}_i$,
where, usually, $\underline{n}_i= (n_{i1},\ldots,n_{iN})$
consists\footnote{Examples are known (see, e.g., Ref.~\cite{Tar}) where
  for master
  integrals some of the indices are equal to two, and not only
  to zero and one. However, in these situation, one can switch
  to another set of master integrals corresponding to indices equal to
  zero or a negative integer value.}
of $n_{ia}=0$, or $1$, or a negative integer value.
Having in mind the experience collected in the process of solving
recurrence relations for various classes of Feynman integrals,
one may hope that, for any given class of Feynman integrals,
the recursion problem can be solved, {\it i.e.}, any Feynman integral
$F(\underline{n})$, with given values of its indices,
can be represented as a finite linear combination
\begin{eqnarray}
  F(\underline{n})&=&\sum_{i\ge1} c_i(\underline{n}) F(\underline{n}_i)\,,
  \label{eqcc}
\end{eqnarray}
with the normalization condition
\begin{eqnarray}
  c_i(\underline{n}_j)&=&\delta_{ij} \,.
  \label{eqcc1}
\end{eqnarray}
The integrals $F(\underline{n}_i)$ are called {\em master}, {\em basic}
or {\em  irreducible} integrals.
Our experience tells us that the coefficient functions
$c_i(\underline{n})$ turn out to be rational functions of the
dimension $d$, the masses and the kinematical invariants
build from the external momenta, so that
the non-trivial
analytical dependence of the Feynman integrals on the dimensionful
quantities
and $d$ is completely contained in the master integrals.

To solve the problem of the reduction it is necessary to identify
the master integrals and to construct an algorithm which allows to calculate
the coefficients $c_i(\underline{n})$.
In concrete situations, the realization of the reduction procedure
to a set of master integrals
turns out to be far from straightforward.
Examples of recent attempts to construct systematic procedures
for solving IBP recurrence relations can be found
in~\cite{Tar,Laporta:2001dd,Kot,GR}.

An alternative approach to solve recurrence relations in a systematic way
was developed in~\cite{Bai1}.
This approach is based on an appropriate
integral representation for
the coefficient functions $c_i(\underline{n})$.
In the case of vacuum Feynman integrals, this representation takes
the form~\cite{Bai1}
\begin{eqnarray}
  c_i(\underline{n}) &\sim&
  \int\ldots\int \frac{\dd x_1 \ldots \dd x_N}{x_1^{n_1} \ldots
    x_N^{n_N}} \left[P(\underline{x})\right]^{(d-h-1)/2} \,,
  \label{eqsol}
\end{eqnarray}
where the parametric integrals have to fulfill
the essential condition that IBP can be applied.
As one can see in the examples discussed in Section~\ref{sec:ex},
the integrals in Eq.~(\ref{eqsol}) can be either performed as
closed contour integrals in the complex plane,
or as iterated integrations over real one-dimensional domains. The latter is
typically between roots of a quadratic polynomial.
Furthermore,
if the singularity of the  $x_i$-integration is given by the
factor $x_i^{-n_i}$ ($n_i=1,2,\ldots$)
it is unreasonable to choose this point as
a boundary of a segment-integration as the corresponding
divergence is not regularized.
However, such a choice turns out to be possible
if (probably,
after some intermediate integrations) an additional
factor of the type $x_i^{-k\ep}$, where $k$ is integer,
appears.

The basic polynomial in Eq.~(\ref{eqsol}) is given by~\cite{Bai1}
\begin{eqnarray}
  P(\underline{x})&=&
  \det\left(\sum_{a=1}^N \tilde{A}^{ij}_a \, (x_a+m_a) \right) \,,
  \label{basic}
\end{eqnarray}
where the determinant is taken with respect to the indices
$i$ and $j$. Thus, $P$ is a homogeneous polynomial of degree $h$
in the variables $x_i$ and the masses.
The matrix $\tilde{A}^{ij}_a$ in Eq.~(\ref{basic}) is defined as follows:
$A^{ij}_a$ as introduced in Eq.~(\ref{denom})
is defined for $i\geq j$. Let us then consider
the quadratic $N\times N$ matrix $A$ (with $N=h (h+1)/2)$, where the
first index is labeled by pairs $(i,j)$ with $i\geq j$, and the second
index is $a$. The corresponding inverse matrix $(A^{-1})^{ij}_a$
(where again $i\geq j$) satisfies
\begin{eqnarray}
  \sum_{a=1}^N  A^{ij}_a  \, (A^{-1})^{i'j'}_a &=& \dl_{ii'}  \dl_{jj'} \,.
\end{eqnarray}
Now $\tilde{A}^{ij}_a$ is defined for all $i$ and $j$ as the symmetrical
extension of $(A^{-1})^{ij}_a$.

As was demonstrated in~\cite{Bai1}, the representation~(\ref{eqsol})
satisfies Eq.~(\ref{eqrr}),
provided one can use IBP in the parametric integrals.
We should remark that the overall normalization in Eq.~(\ref{eqsol})
is not fixed in advance but is
adjusted after the construction of the
coefficient functions, as a virtue of Eq.~(\ref{eqcc1}). For the same
reason, the basic polynomial is determined up to a
factor which is independent of the variables $x_i$.

General Feynman integrals can be
reduced to the vacuum case~\cite{Bai1,Bai7}. In case an external
momentum is reduced to a mass shell, $p_i^2=m_i^2$ (where the mass
$m_i$ is one of the internal masses of the diagram), one includes
into the procedure all the terms of the formal Taylor expansion in
$p_i^2$ at $p_i^2=m_i^2$. In the case of a general external
momentum squared one can still consider similar terms of the
Taylor expansion in $p_i^2-M_i^2$ where, this time, $M_i$ is not
equal to some of the internal masses. This Taylor
expansion is indeed a Taylor series (rather than a formal series)
because the point $p_i^2=M_i^2$ is not singular. In this
situation, one is usually interested only in the value at
$p_i^2=M_i^2$ and not in the derivatives at this
general point. Thus for the corresponding Feynman integrals one
considers only the value $n_i=1$ of the corresponding index $n_i$.
The transition to a vacuum problem effectively increases the
number of loops which enters the exponent of the basic polynomial
$P$ in Eq.~(\ref{eqsol}) where the loop number $h$ refers to the
corresponding vacuum Feynman graph. For example, in a recursive
problem for a class of propagator diagrams, the transition to the
corresponding vacuum problem, which is obtained after gluing the
external lines, increases the loop number by one.
We would like to stress that this translation to the level of
vacuum Feynman integrals is not necessarily accompanied
by a corresponding vacuum graph.
However, in the language of Feynman integrals defined through
Eq.~(\ref{eqbn}) such a description is always possible.

It should be mentioned that although Eq.~(\ref{eqsol}) has been
applied to solve IBP recurrence relations in some physical
situations~\cite{Bai4,Bai8}
no instructions how to apply this representation
are available in the literature.

The goal of our paper is twofold.
First we shall describe how the parametric representation of
Eq.~(\ref{eqsol}) can be
used to characterize the master integrals and
to construct an algorithm for evaluating the coefficients of these
master integrals.
Second, we shall choose several typical examples to exemplify the method.
As a non-trivial two-loop example, we
discuss in detail the class of diagrams appearing in the
calculation of the two-loop heavy quark potential.
The main motivation for this is that in Refs.~\cite{Pet,Sch}
no complete procedure has been presented but only
the necessary Feynman integrals were calculated.
A reduction procedure is described in the thesis~\cite{Sch-thesis}
which is based on IBP accompanied by the method developed in~\cite{Tar}.
In Ref.~\cite{KPSS1}, where $1/(m_q r^2)$ corrections
to the potential were evaluated,
a standard approach to solve the IBP relations has been used
and an algorithm has been developed to evaluate the adequate
subclass of the diagrams.
In this paper we present a complete procedure for evaluating a
general diagram of this class which is certainly useful if higher
dimensional operators are considered within the framework of
non-relativistic
QCD.

The outline of this paper is as follows:
in the next section, we discuss the identification of the master
integrals and present general prescriptions for applying
Eq.~(\ref{eqsol}).
In Section~\ref{sec:ex}, we discuss the one- and two-loop examples
and, finally, in Section~\ref{sec:concl} we present our conclusions.

%%%%%%%%%%%%%%%%%%%%%%%%%%%%%%%%%%%%%%%%%%%%%%%%%%%%%%%%%%%%

\section{\label{sec:method}Classification
of master integrals and construction of coefficient functions}

In this section we present practical
prescriptions to perform a reduction procedure with the help of
Eq.~(\ref{eqsol}).
We shall assume that any Feynman integral can be
decomposed according to Eqs.~(\ref{eqcc}) and~(\ref{eqcc1}).
The method decomposes into two parts: the classification of the master
integrals and the construction of the coefficient functions.

\subsection{\label{sub:id}Identifying master integrals}

Let us in a first step consider the integrals where the
indices corresponding to irreducible numerators
are set to zero. The other indices we allow to be either zero or one.
In the following we nullify all scaleless integrals
(e.g., massless tadpoles) which is natural within
the framework of dimensional regularization.

Before analyzing the remaining non-zero integrals
with the indices one and zero, let us
remember that the goal of every known reduction is to reduce some
of the indices to zero. Indeed, experience tells us that a master
integral $F(\un_i)$ never appears in the reduction of a given
Feynman integral $F(\un)$ if $n_j\leq 0$ and  $n_{ij}>0$.
For this reason we adopt the following natural
condition for the coefficient
function $c_i(\un)$ of $F(\un_i)$:
\begin{eqnarray}
  c_i(\un) &= &0 \quad\mbox{if}\quad n_j\leq 0
  \quad\mbox{and}\quad n_{ij}=1 \quad\mbox{for}\quad
 j=1,\ldots,N
 \,.
\end{eqnarray}
This condition is easily realized in an automatic way
by choosing the integration domain for $x_j$ in Eq.~(\ref{eqsol})
as a closed contour around the origin in the complex $x_j$-plane, which
from now on will always be implied.
As a consequence,
the integration over the variables $x_j$
in Eq.~(\ref{eqsol}),
where $n_{ij}=1$ in the corresponding master integral,
is performed with the help of the Cauchy
theorem and thus reduces to Taylor expansions of order $n_j-1$
of the integrand in $x_j$.
Eventually, Eq.~(\ref{eqsol}) results in a sum of
terms of the form
\begin{eqnarray}
  \int\ldots\int \left[P_i(\ux)\right]^{z-n_d} \,
  \prod_{j:n_{ij}\leq 0} \frac{\dd x_j}{x_j^{n^\prime_j}}
  \,,
\label{eqsol1}
\end{eqnarray}
where $z=(d-h-1)/2$,  $n_d$ is a non-negative integer,
$n^\prime_j$ are some integer exponents, and
$P_i(\ux)$ is obtained from $P(\ux)$ after setting to zero all
variables $x_j$ where $n_{ij}=1$.

Let us now analyze the remaining candidates in view of being master
integrals
with the help of the representation~(\ref{eqsol1}).
If $P_i=0$ for a given candidate  $F(\un_i)$ ({\it i.e.}, for $\un_i$
with $n_{ij}=1$ or $0$) it is naturally to put the corresponding
coefficient function to zero. In other words, such a Feynman
integral is recognized as a reducible integral in the sense of the
reduction problem. This is the simplest condition of
reducibility. A more general condition is that any resulting integral
of Eq.~(\ref{eqsol1}) involves an integral without scale. In this
situation it is natural to prescribe a zero value for
it\footnote{At this point we once again
  apply a commonly accepted agreement within dimensional
  regularization to put any scaleless integral to zero.
  Observe that such a prescription, in an extended form, is also
successfully
  applied within the strategy of expansion by regions~\cite{BS} when
expanding Feynman integral in various limits of masses and momenta.
Moreover scaleless integrals that are not dimensionally regularized
(e.g., connected with the contribution of the potential region)
are also consistently put to zero with this strategy.}
and, therefore, consider the given Feynman integral
$F(\un_i)$ as reducible in the sense of the reduction problem
and the corresponding decomposition given in Eq.~(\ref{eqcc}).
It can happen that a scaleless integral in Eq.~(\ref{eqsol})
can either be seen immediately, {\it i.e.},
an integral over a ``pure'' power of
some variable arises,
or only after some intermediate integrations.

After trying all possible combinations of zero and one we obtain a
list of master integrals, which fixes the terms on the
right-hand side of Eq.~(\ref{eqcc}).

However, this is not all. Sometimes, in addition to a given
master integral $F(\un_i)$ with $\un_i$ consisting of $n_{ij}=1$ or
$n_{ij}=0$, one has to consider master integrals which
differ from $F(\un_i)$ by some indices $n_{ij}<0$.
The number of such additional master integrals
is dictated by the degree of the
polynomial $P_i$ with respect to some of the parameters $x_j$.
In practice it turns out to be advantageous to complete the
list of the master integrals in the second step where
algorithms for the computation of the coefficient functions $c_i(\un)$ are
constructed.

Sometimes it is obvious that a given candidate is
reducible due to other arguments like the simple application of the
triangle rule. However, it turns
out to be rather convenient to see this property directly by analyzing the
corresponding ``reduced'' polynomial $P_i$,
as this provides a useful additional check.

%%%%%%%%%%%%%%%%%%%%%%%%%%%%%%

\subsection{Constructing coefficient functions}

The set of the master integrals with the indices zero and one as obtained in
the previous Subsection is partially ordered in the natural way.
We write $F(\un_1)< F(\un_2)$ if $n_{1j} \leq n_{2j}$ for all $j$
and the strict inequality holds at least for one index.
In this way all master integrals can be grouped into so-called families
which are by definition disjoint. As we will see below this ordering is
crucial in the construction of the coefficient functions.

Let us start from the master integrals
which have the most non-zero indices where only indices corresponding to
the propagators but not to the numerators are counted.
These integrals are maximal in the sense of the hierarchy defined above.
For a maximal master integral $F(\un_i)$, the corresponding
coefficient function $c_i(\un)$ in Eq.~(\ref{eqcc})
is replaced by the corresponding parametric integral~(\ref{eqsol})
which results in a sum of terms as given in Eq.~(\ref{eqsol1})
according to our
agreement of choosing integration contours for the variables
associated with the indices equal to one.
We want to stress that only those integrations over $x_i$ are
non-trivial where the corresponding index of the master integral is zero.
Experience shows that these remaining integration can be performed in terms
of
gamma functions, for the general integer values of the
indices.\footnote{Typically,
  integrals involving the Euler beta function arise as we will see in the
  examples discussed in Section~\ref{sec:ex}.}
At this point one applies Eq.~(\ref{eqcc1}) to normalize the resulting
expression.

Let us now suppose that we are dealing with a master integral
which is not maximal, {\it i.e.},
we have two master integrals which fulfill the
hierarchy $F(\un_2)< F(\un_1)$.
This means  that  if $n_{2i}=1$ we have also $n_{1i}=1$.
To construct an algorithm for the coefficient function
$c_2(\un)$ we start with the case of negative indices $n_j$
for those indices $j$ where $n_{1i}=1$ as in this case we
have $c_1(\un)=0$.
Experience shows that the integrations for $c_2(\underline{n})$
result in ratios of gamma functions which in particular can be used to
obtain $c_2(\un_2)=1$.
In a next step one considers the case $n_j>0$.
Then the corresponding parametric representation~(\ref{eqsol})
usually leads to integrals (cf. Eq.~(\ref{eqsol1}))
which cannot be evaluated in gamma
functions.\footnote{However, sometimes this is possible as we will see
  in the examples below.}
Thus at first sight
it looks hopeless to achieve that the coefficient functions have to be
rational functions of $d$ (and eventually other dimensionful quantities).
The way out of this problem is as follows:
actually we have to look for an expression for the coefficient function
$c_2(\underline{n})$ which is a linear combination of
$c_1(\un)$ and the representation~(\ref{eqsol})
for $c_2(\un)$. Denoting the latter by $c_2^0(\un)$ one has
\begin{eqnarray}
  c_2(\un) &=& c_2^0(\un) + A\,c_1(\un)
  \,,
  \label{eq:c2c1}
\end{eqnarray}
where the constant $A$ is determined by the condition $c_2(\un_1)=0$
(cf. Eq.~(\ref{eqcc1}))
and is thus given by
\begin{eqnarray}
  A &=& - c_2^0(\un_1)
  \,.
\end{eqnarray}

At this stage it turns out that it is advantageous to use
IBP applied to the parametric integrals of Eq.~(\ref{eqsol1}).
Note that the complexity is significantly reduced as compared to the
original
integral simply because there are much less variables involved.
The basic idea is to express any given parametric integral in
terms of auxiliary (parametric)  master integrals
and expressions which are straightforwardly evaluated in gamma functions.
The dependence on the new auxiliary master integrals has to drop
out in order to guarantee that
the coefficient functions are rational
functions of $d$ (and eventually the other dimensionful
parameters).
Such a cancellation turns to be a good check of the algorithm.
This is similar to the cancellation of spurious poles when applying general
prescriptions for asymptotic expansion of Feynman integrals in
various limits of momenta and masses~\cite{BS}.
In our particular example of a hierarchy of two master integrals
one presumably only needs one auxiliary parametric master
integral.

For a generic tower of master integrals
the construction and cancellation of the auxiliary master integrals goes
along the same lines.
When constructing the corresponding coefficient function
one uses a linear combination of the coefficient under consideration and
all coefficients of the higher master integrals.
The constants are determined with the help of Eq.~(\ref{eqcc1}).
Explicit examples can be found in Section~\ref{sec:ex}.

Let us stress again that in general it is rather simple to compute the
coefficient function of a maximal master integral as it has
many indices equal to one. On the other hand, following the procedure
described above, it is possible to either reduce
the indices of the coefficient of a non-maximal master integral
in order to make an evaluation in $\Gamma$ functions possible,
or relate it to the coefficient functions of higher master integrals.
This is achied with the help of IBP applied to the parametric integrals
over the $x-$parameters.

At this point it may happen that the initial list
of the master integrals obtained in Section~\ref{sub:id}
is not sufficient and has to be extended by
additional master integrals where some of the indices are
negative. This necessity is connected to the type of
solution of the recursive problem for the auxiliary parametric integrals
and thus to the basic polynomial
$P_i(\ux)$.

We consider the construction of a coefficient functions as completed if for
given indices an algorithm  (usually realized on a
computer) for its evaluation is provided.
However, in simple situations, one can even
obtain explicit expressions for
the coefficients as functions of the indices.

After we have obtained the list of the master integrals it is reasonable to
check whether there are identical expressions among them, although
they all have appeared in a different way. If some of the master
integrals within one family are equal simply because the corresponding
integrals are identical, e.g., due to some symmetry, it is natural
to consider one master integral instead. The corresponding coefficient
function is the sum of the coefficient functions of the original
integrals. From the mathematical point of view,
the best solution would be to achieve the absolute minimal number of the
master integrals in Eq.~(\ref{eqcc}). However, for practical
purposes, this is not very important.

Finally let us note that it is possible to identify an
integral as master and then immediately construct an algorithm for
the corresponding coefficient function, rather than strictly
decompose the procedure into the two steps as described above.

%%%%%%%%%%%%%%%%%%%%%%%%%%%%%%%%%%%%%%%%%%%%%%%%%%%%%%%%%%%%

\section{\label{sec:ex}One- and two-loop examples}

\begin{figure}[t]
  \begin{center}
      \leavevmode
      \epsfxsize=16cm
      \epsffile[90 400 500 470]{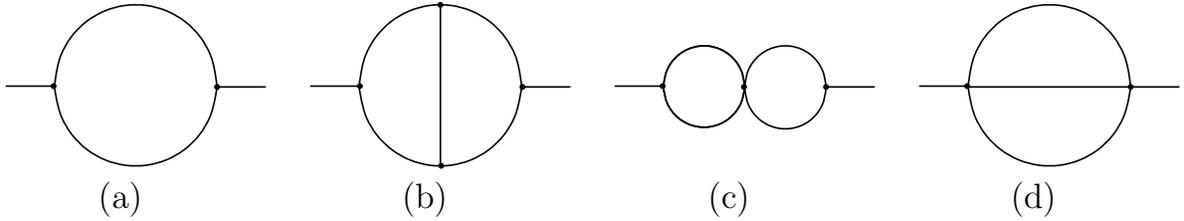}
  \end{center}
  \caption{\label{fig:prop}
    Feynman diagrams for one- and two-loop massless propagator ((a) and
(b)).
    At two-loop order there are two master integrals which are shown
    in (c) and (d).
          }
\end{figure}

\subsection{One-loop propagator diagram}

Let us start with the simple example of a massless one-loop diagram
with one external momentum (a so-called propagator diagram) which is
given by (see also Fig.~\ref{fig:prop}(a))
\begin{eqnarray}
  F(n_1,n_2) &=&
  \int \frac{{\rm d}^dk}{(k^2)^{n_1} [(k-q)^2]^{n_2}}\,,
  \label{eq:1lprop}
\end{eqnarray}
where the usual prescriptions like $k^2=k^2+i 0$, etc. are implied.
According to \cite{Bai1}, the transition to the corresponding vacuum
problem reduces to
adding a new propagator, $1/(q^2-s)$, with some mass squared $s$.
Furthermore one has $h=2$.
We want to consider the value of our diagram at some general point
and are not interested in higher terms of the Taylor expansion in $q^2$.
Therefore we consider only the index $n_3=1$
corresponding to the additional propagator.
The integration contour for the corresponding variable $x_3$ is
taken as a Cauchy contour around the origin. Thus we write down the
corresponding basic polynomial (\ref{basic}) and reduce it by
setting $x_3=0$.
After that the resulting basic polynomial reads
\begin{eqnarray}
  P(x_1,x_2) &=& (q^2)^2  - 2 q^2 (x_1+x_2)+ (x_1-x_2)^2
  \,.
  \label{eq:1lprop_pol}
\end{eqnarray}
The only possible candidate for a
master integral is  $F(1,1)$ because we obtain
massless tadpoles when we put one of the indices to zero.
We have
\begin{eqnarray}
  I_1 &=& F(1,1) \,\,=\,\,
 i\pi^{d/2} (-q^2)^{d/2-2}\frac{\Gamma(2-d/2)\Gamma^2(d/2-1)}
  {\Gamma(d-2)}
  \,.
\end{eqnarray}
The corresponding coefficient function is easily obtained from
Eq.~(\ref{eqsol}) with the help of the Cauchy theorem
\begin{eqnarray}
  c_1(n_1,n_2) = N_1
  \frac{1}{(n_1-1)!}\left(
  \frac{\partial}{\partial x_1}\right)^{n_1-1}
  \frac{1}{(n_2-1)!}\left(
  \frac{\partial}{\partial x_2}\right)^{n_2-1}
  \left[P(x_1,x_2)\right]^{(d-3)/2}\Bigg|_{x_i=0}
  \,,
  \label{eq:1lprop_coef}
\end{eqnarray}
where $N_1$ is a ($d$-dependent) normalization factor introduced in order to
fulfill Eq.~(\ref{eqcc1}). For our choice of $P(x_1,x_2)$ it reads
\begin{eqnarray}
  N_1 &=& \left(q^2\right)^{(d-3)}
  \,.
\end{eqnarray}

It is straightforward to implement
Eq.~(\ref{eq:1lprop_coef}) on a computer.
The result agrees with the known one which in this case, of
course, is trivially obtained by a straightforward Feynman
parameterization.

%%%%%%%%%%%%%%%%%%%%%%%%%%%%%%

\subsection{\label{sec:prop2l}Two-loop propagator diagram}

The situation becomes slightly more complicated at two-loop level where the
Feynman integral is given by (cf. Fig.~\ref{fig:prop}(b))
\begin{eqnarray}
  F(\un) &=&
  \int \frac{{\rm d}^dk}{
    (k^2)^{n_1} (l^2)^{n_2} [(k-q)^2]^{n_3} [(l-q)^2]^{n_4} [(k-l)^2]^{n_5}
    }
  \,.
  \label{eq:2lprop}
\end{eqnarray}
The transition to vacuum integrals is performed in analogy to the
previous one-loop case by introducing one more propagator
connected with the external invariant $q^2$
and $h=3$ (see also the next example for more details). 
As before, we set the
corresponding $x$-parameter to zero and obtain the following
basic polynomial
\begin{eqnarray}
  P(x_1,\ldots,x_5) &=&
  -x_1 x_2 x_3 + x_2^2 x_3 + x_2 x_3^2
  + x_1^2 x_4 - x_1 x_2 x_4 - x_1 x_3 x_4 -
  x_2 x_3 x_4 + x_1 x_4^2
  \nonumber\\&&\mbox{}
  + x_1 x_2 x_5 - x_2 x_3 x_5 - x_1 x_4 x_5 + x_3 x_4 x_5
  + q^2 [ -x_1 x_3 + x_2 x_3 + x_1 x_4
  \nonumber\\&&\mbox{}
  - x_2 x_4 + x_1 x_5 + x_2 x_5 +
  x_3 x_5 + x_4 x_5 - x_5^2 ]
  + (q^2)^2 x_5 \;.
\end{eqnarray}
An analysis of the integrals with indices equal to zero leads to
two master integrals shown in Fig.~\ref{fig:prop}(c) and~(d)
\begin{eqnarray}
  I_1 &=& F(1,1,1,1,0) \,\,=\,\,
(i\pi^{d/2})^2  (-q^2)^{d-4}\frac{\Gamma^2(2 - d/2) \Gamma^4(d/2-1)}
  {\Gamma^2(d-2)}
  \,,\nonumber\\
  I_2 &=& F(0,1,1,0,1)=F(1,0,0,1,1) \,\,=\,\,
 - (i\pi^{d/2})^2 (-q^2)^{d-3} \frac{\Gamma(3 - d) \Gamma^3(d/2-1)}
  {\Gamma(3d/2-3)}
  \,,
\end{eqnarray}
where the second one appears in two ways.
As far as the coefficient functions are concerned the
integration over those $x_i$ with index ``1'' in the corresponding
master integral is trivially obtained through differentiation of $P(\ux)$.
In the case of $c_1$ this leads to the following remaining integral
\begin{eqnarray}
  I_h^{(1)}(\alpha,\beta) \,\,=\,\,
  \int_0^{q^2} {\rm d}x_5 \, x_5^\alpha (q^2-x_5)^\beta &=&
  \left(q^2\right)^{\alpha+\beta+1} \frac{\Gamma(\alpha+1)\Gamma(\beta+1)}
  {\Gamma(\alpha+\beta+2)}
  \,.
  \label{eq:Ih1}
\end{eqnarray}
Here and in the following
$\alpha$ and $\beta$ depend on the dimension $d$ whereas $k$ represent
integer
indices.
On the other hand, for $c_2$ one has to solve a two-dimensional
integral over $x_1$ and $x_4$ which is calculated
as follows
\begin{eqnarray}
  I_h^{(2)}(\alpha_1,\alpha_4,\beta) &=&
  \int_0^\infty {\rm d}x_1 \, {\rm d}x_4 \,
  x_1^{\alpha_1} x_4^{\alpha_4} (q^2+x_1+x_4)^\beta
  \nonumber\\
  &=&
  \left(q^2\right)^{\alpha_1+\alpha_4+\beta+2}
  \frac{\Gamma(\alpha_1+1)\Gamma(\alpha_4+1)
    \Gamma(-\alpha_1-\alpha_4-\beta-2)}
  {\Gamma(-\beta)}
  \,.
  \label{eq:Ih2}
\end{eqnarray}
Again a small computer program takes over the differentiation and
integration parts and finally
results to the two-loop integral
\begin{eqnarray}
  F(n_1,n_2,n_3,n_4,n_5) &=&
  c_1(n_1,n_2,n_3,n_4,n_5) I_1
  \nonumber\\&&\mbox{}
  + \left[ c_2(n_1,n_2,n_3,n_4,n_5) + c_2(n_2,n_1,n_4,n_3,n_5) \right] I_2
  \,,
\end{eqnarray}
in agreement with the usual approach based on IBP~\cite{IBP}.

%%%%%%%%%%%%%%%%%%%%%%%%%%%%%%

\begin{figure}[t]
  \begin{center}
      \leavevmode
      \epsfxsize=16cm
      \epsffile[90 400 500 470]{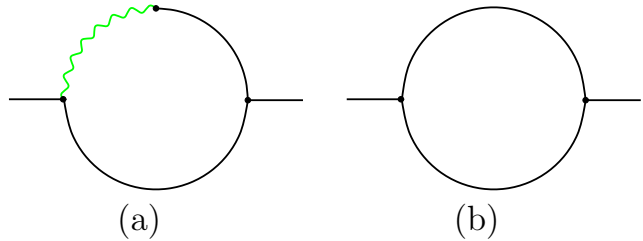}
  \end{center}
  \caption{\label{fig:pot1}
    (a) Generic Feynman diagram needed for the one-loop
    heavy quark potential. There two master integrals: one corresponds to
    the
    very diagram with all indices equal to one and the second one is
    shown~(b).
    }
\end{figure}

%%%%%%%%%%%%%%%%%%%%%%%%%%%%%%

\subsection{One-loop diagram for the heavy quark potential}

The integrals studied in this and the next
subsection are useful in the context of
non-relativistic QCD (NRQCD) where in addition to relativistic
propagators (for gluons and light quarks) also non-relativistic ones
(for heavy quarks) appear.
The general one-loop integral is shown in Fig.~\ref{fig:pot1} and has the
form
\begin{eqnarray}
  F(n_1,n_2,n_3) &=&
  \int \frac{{\rm d}^dk}{(k^2)^{n_1}
  [(k-q)^2]^{n_2} (v \cdo k)^{n_3}}\,,
  \label{1lp}
  \label{eq:F1lp}
\end{eqnarray}
with $v \cdo k$ understood as $v \cdo k+i 0$.
Here $q$ is the external momentum and
$v$ is a constant vector which fulfills the condition $v \cdo q =0$.

When turning to the corresponding vacuum problem we consider,
in addition to $k^2$, $q \cdo k$ and  $v \cdo k$,
three extra kinematical invariants, $q^2, v \cdo q, v^2$,
which play the role of inverse auxiliary propagators.
The latter are obtained after interpreting the expression in
Eq.~(\ref{eq:F1lp}) as a three-point function with external momenta $q$ and
$v$ and closing the three external lines in a new vertex.
This formally leads to a three-loop vacuum problem and thus to $h=3$.

In the notation of Eq.~(\ref{eqbn}) we have $D_1 = k^2, D_2 = (k - q)^2,
D_3 = k\cdo v + v^2, D_4 = v^2, D_5 = q^2$ and $D_6 = (q + v)^2$
which defines the matrix $A$ according to Eq.~(\ref{denom}).
The symmetrical extension of the corresponding inverse matrix leads to
(after identifying the six invariants with $x_1,\ldots,x_6$)
\begin{eqnarray}
  \left(
  \begin{array}{ccc}
    x_1 & (x_1 - x_2 + x_5)/2 & x_3 - x_4
    \\
    (x_1 - x_2 + x_5)/2 & x_5 & (-x_4 - x_5 + x_6)/2 
    \\
    x_3 - x_4 & (-x_4 - x_5 + x_6)/2 & x_4
  \end{array}
  \right)\,.
\end{eqnarray}
The basic polynomial $P$ is obtained from the determinant of this matrix
with the help of Eq.~(\ref{basic}).
Moreover, we shift the variables $x_i$ by the corresponding
effective masses:
$x_3 \to x_3 + v^2, x_4  \to x_4 + v^2, x_5  \to x_5 + q^2,
x_6  \to x_6 + (q + v)^2$.
Note, that as in the previous examples we
are not interested in higher order Taylor coefficients of the additional
kinematical invariants. Thus we can reduce the basic polynomial by setting
$x_4=x_5=x_6=0$ and finally obtain
\begin{eqnarray}
  P(x_1,x_2,x_3) &=&
  (q^2)^2 v^2 + v^2 \left(x_1 - x_2\right)^2
  + 2 q^2 \left[v^2 \left(x_1 + x_2\right) - 2 x_3^2\right]
  \,,
\end{eqnarray}
which leads to the two master integrals
\begin{eqnarray}
  I_1 &=& F(1,1,1) \,\,=\,\,
  - i\pi^{d/2} \frac{(-q^2)^{d/2-5/2}\sqrt{\pi} }{v}
  \frac{\Gamma(5/2-d/2)\Gamma^2(d/2-3/2)}{\Gamma(d-3)}
  \,,\nonumber\\
  I_2 &=& F(1,1,0) \,\,=\,\,
  i\pi^{d/2} (-q^2)^{d/2-2}\frac{\Gamma(2-d/2)\Gamma^2(d/2-1)}
  {\Gamma(d-2)}
  \,.
\end{eqnarray}

Since for $I_1$ all indices are one
the coefficient function $c_1$ is obtained in analogy to
the one of Eq.~(\ref{eq:1lprop_coef}).
In the case of $c_2$ the indices $n_1$ and
$n_2$ are treated with the help of differentiations which leads to the
following type of integral for $x_3$
\begin{eqnarray}
  I_h^{(3)}(k_3,\alpha) &=& \int_{-a}^{a}
  {\rm d}x_3 \, x_3^{k_3} \left( a^2 - x_3^2 \right)^\alpha
  \,.
  \label{eq:Ih3}
\end{eqnarray}
Here $k_3$ is an integer but $\alpha$ depends on $d$.
$I_h^{(3)}(k,\alpha)$ can be interpreted as a principal value integral
leading to
\begin{eqnarray}
  I_h^{(3)}(k,\alpha) &=& \left\{
    \begin{array}{ll}
      \displaystyle
      \left(a^2\right)^{\alpha+k/2+1/2}
      \frac{\Gamma(k/2 + 1/2) \Gamma(\alpha+1)}
      {\Gamma(\alpha + k/2 + 3/2)}
      \qquad
      & \mbox{for $k$ even}
      \\
      \displaystyle
      0 & \mbox{for $k$ odd}
    \end{array}
  \right.
  \,.
  \label{eq:Ih3a}
\end{eqnarray}

The final result
\begin{eqnarray}
  F(n_1,n_2,n_3) &=& c_1(n_1,n_2,n_3) I_1 + c_2(n_1,n_2,n_3) I_2
  \,,
\end{eqnarray}
agrees with the explicit analytical result (see, e.g.,
Ref.~\cite{Sch-thesis}).

%%%%%%%%%%%%%%%%%%%%%%%%%%%%%%

\begin{figure}[t]
  \begin{center}
      \leavevmode
      \epsfxsize=16cm
      \epsffile[80 400 500 460]{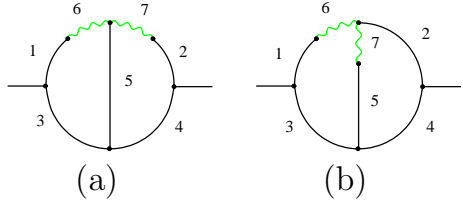}
  \end{center}
  \caption{\label{fig:potab}
    Feynman diagrams corresponding to case~A (a) and case~B (b).
          }
\end{figure}

%%%%%%%%%%%%%%%%%%%%%%%%%%%%%%

\subsection{Two-loop diagrams for the heavy quark potential}

At two-loop order two classes of Feynman integrals are needed which
we will refer to as case~A and case~B
\begin{eqnarray}
  F_A(\un) &=&
  \int\int\frac{{\rm d}^dk{\rm d}^dl}{(k^2)^{n_1}(l^2)^{n_2}
    [(k-q)^2]^{n_3}[(l-q)^2]^{n_4}[(k-l)^2]^{n_5}(v \cdo k)^{n_6}(v \cdo
    l)^{n_7}}
  \,,
  \label{2lpA}
  \\
  F_B(\un) &=&
  \int\int\frac{{\rm d}^dk{\rm d}^dl}{(k^2)^{n_1}(l^2)^{n_2}
    [(k-q)^2]^{n_3}[(l-q)^2]^{n_4}[(k-l)^2]^{n_5}(v \cdo k)^{n_6}
    [v \cdo (k-l)]^{n_7} }
  \,.
  \label{2lpB}
\end{eqnarray}
They are shown in Fig.~\ref{fig:potab}.
The corresponding basic polynomials read
\begin{eqnarray}
  P_A(x_1,\ldots,x_7) &=&
  - [ x_2 x_6 - x_4 x_6 + (-x_1 + x_3) x_7 ]^2
  + v^2 \{x_1^2 x_4 + x_3 (x_2^2 + x_2 (x_3 - x_4 - x_5)
  \nonumber\\&&\mbox{}
  + x_4 x_5) -
  x_1 [ x_2 (x_3 + x_4 - x_5) + x_4 (x_3 - x_4 + x_5) ] \}
  \nonumber\\&&\mbox{}
  + (q^2)^2 [ v^2 x_5 - (x_6 - x_7)^2 ]
  + q^2 \{
  v^2 [ (x_3 + x_4 - x_5) x_5 + x_2 (x_3 - x_4 + x_5)
  \nonumber\\&&\mbox{}
  + x_1 (-x_3 + x_4 + x_5) ] +
  2 [ x_2 x_6 (-x_6 + x_7) + x_4 x_6 (-x_6 + x_7)
  \nonumber\\&&\mbox{}
  + x_7 (x_1 x_6 + x_3 x_6 - 2 x_5 x_6 - x_1 x_7 - x_3 x_7) ] \}
  \,,
  \\
  P_B(x_1,\ldots,x_7) &=& P_A(x_1,x_2,x_3,x_4,x_5,x_6,x_6-x_7)
  \label{eq:polb}
  \,.
\end{eqnarray}

\begin{figure}[t]
  \begin{center}
      \leavevmode
      \epsfxsize=16cm
      \epsffile[80 320 500 520]{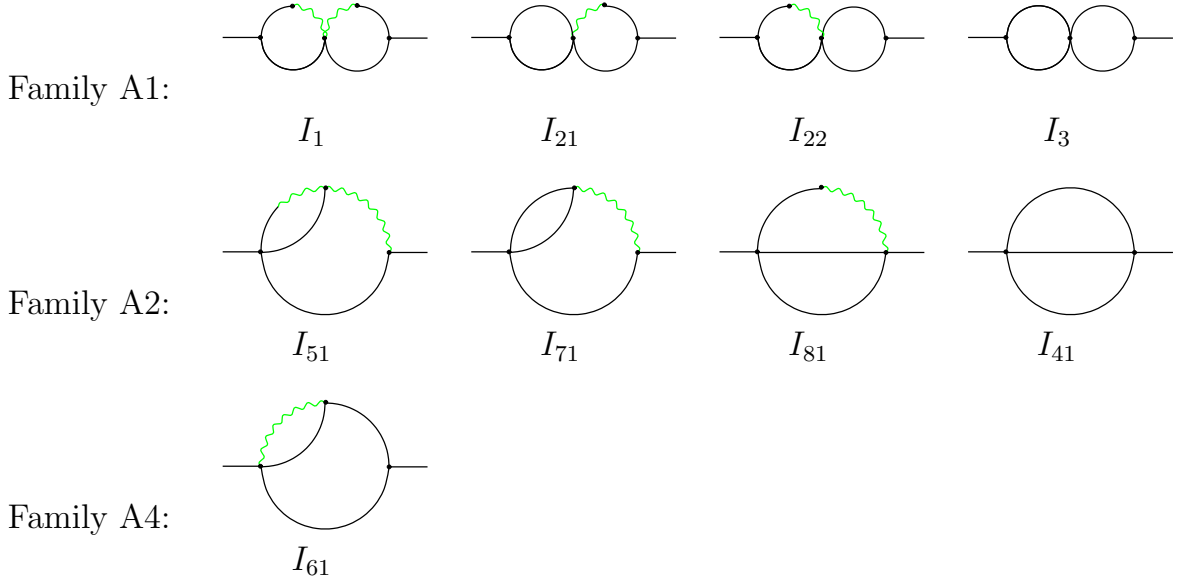}
  \end{center}
  \caption{\label{fig:pota}
    Feynman diagrams corresponding to the master integrals
    of case~A. In addition to $I_{61}$ there is also a master integral
    ($I_{61a}$) containing an irreducible numerator (see text).
          }
\end{figure}

The application of the procedures described in Section~\ref{sec:method}
to case~A leads to the following families of master integrals which
are also pictured in Fig.~\ref{fig:pota}.
As far as the notation is concerned the first index labels the different
master integrals. In case the master integrals are equal we introduce a
second
index for further specification. An additional subscript ``$a$'' is
added for those master integrals where one index is ``$-1$''.

\begin{itemize}
\item Family~A1.
  One has four master integrals which obey the following hierarchy:
  $I_1 > \{I_{21},I_{22}\} > I_3$ with
  \begin{eqnarray}
    I_1 &=& F_A(1,1,1,1,0,1,1)\,,\nonumber\\
    I_{21} &=& F_A(1,1,1,1,0,0,1)\,,\nonumber\\
    I_{22} &=& F_A(1,1,1,1,0,1,0)\,,\nonumber\\
    I_3 &=& F_A(1,1,1,1,0,0,0)\,.
  \end{eqnarray}

\item Family~A2.
  One has four master integrals which obey the following hierarchy:
  $I_{51} > \{I_{71},I_{81}\} > I_{41}$ with
  \begin{eqnarray}
    I_{51} &=& F_A(1,0,0,1,1,1,1)\,,\nonumber\\
    I_{71} &=& F_A(1,0,0,1,1,0,1)\,,\nonumber\\
    I_{81} &=& F_A(1,0,0,1,1,1,0)\,,\nonumber\\
    I_{41} &=& F_A(1,0,0,1,1,0,0)\,.
  \end{eqnarray}

\item Family~A3
  is symmetrical to Family~A2 with respect to
  $1\lra 2, 3\lra 4, 6\lra 7$. It contains the master integrals
  $I_{52}$, $I_{72}$, $I_{82}$ and $I_{42}$.

\item Family~A4
  contains the master integrals
  \begin{eqnarray}
    I_{61} &=& F_A(0,1,0,1,1,1,0)\,,\nonumber\\
    I_{61a} &=& F_A(0,1,0,1,1,1,-1)\,.
  \end{eqnarray}

\item Family~A5
  is symmetrical to Family~A4 with respect to $1\lra 2, 3\lra 4, 6\lra 7$.
  It contains the master integrals $I_{62}$ and $I_{62a}$.

\end{itemize}
As expected, the master integrals of the massless scalar two-loop diagram
discussed in Section~\ref{sec:prop2l} also appear in the above list.

As has already become clear from the examples discussed so far, one expects
the
appearance of complicated expressions for the coefficient functions of the
simple master integrals. Indeed, in the case of
the coefficient function $c_{1}$
six out of seven indices can be treated with the help of differentiations
and the remaining one-dimensional integral is easily solved with the help
of $I_h^{(1)}$ given in Eq.~(\ref{eq:Ih1}).
The situation is similar for $c_{22}$ (and $c_{21}$ which can be obtained by
exploiting the symmetry) where the remaining two-fold integration
over $x_7$ and $x_5$ is performed with the help of
\begin{eqnarray}
  I_h^{(4)}(k,\alpha_1, \alpha_2) &=&
  \int_a^b {\rm d}x\, x^{k} (x - a)^{\alpha_1} (b - x)^{\alpha_2}
  \nonumber\\
  &=&
  \sum_{r=0}^{k} a^{k - r}  (b-a)^{\alpha_1+\alpha_2+r+1}
  \frac{k!}{(k-r)! r!}
  \frac{\Gamma(1 + \alpha_2)\Gamma(1 + \alpha_1+r)}
  {\Gamma(\alpha_1+\alpha_2+r+2)}
  \,.
  \label{eq:Ih4}
\end{eqnarray}
and Eqs.~(\ref{eq:Ih3}) and~(\ref{eq:Ih3a}), respectively.
In principle, instead of performing the $x_5$ integration
one could also introduce an auxiliary master integral
as a result of a two-dimensional ($x_5$ and the dimension $d$) recursion
problem.
This master integral would cancel after considering the
proper linear combination with $c_1$
as described around Eq.~(\ref{eq:c2c1}).

For the coefficient $c_3$ there are three non-trivial integrations
($x_5$, $x_6$, $x_7$) left.
In case one of the indices $n_5$, $n_6$ or $n_7$ is less or equal to zero
one can use various combinations of the auxiliary integrals
$I_h^{(i)}$ $(i=1,\ldots,4)$ listed above. Thereby it is advantageous to
perform the integration corresponding to the negative index first.
If, on the contrary, $n_5$, $n_6$ and $n_7$ are positive an immediate
integration
seems not to be possible. However, from the corresponding three-parametric
integral representation it is simple to derive recurrence relations which
shift at least one of the indices to zero, eventually at the cost of
increasing the dimension. The latter does not constitute a problem since the
whole formulation of our procedure is in $d$ dimensions.
Thus, also in this case the integration can be performed in terms of
$\Gamma$ functions.
In principle one could be forced to introduce three auxiliary master
integrals
and build the proper linear combinations with $c_1$, $c_{21}$ and $c_{22}$.
However, it turns out that the corresponding constants, $c_3(\un_i)$
($i=1,2,3$) are zero.

For the coefficient function $c_{51}$ only two non-trivial
integrations over $x_2$ and $x_3$ are involved which can be performed
with the help of Eq.~(\ref{eq:Ih2}).

The situation is similar for $c_{71}$. It is convenient to perform in
a first step the $x_6$ integration with the help of $I_h^{(3)}$
of Eq.~(\ref{eq:Ih3}). For the remaining integration over $x_2$ and $x_3$
the
formula
\begin{eqnarray}
  I_h^{(5)}(k,\alpha_2,\alpha_3,\beta) &=&
  \int_0^\infty {\rm d}x_2\,{\rm d}x_3\,
  x_2^{\alpha_2} x_3^{\alpha_3} \left(q^2+x_2+x_3\right)^\beta
  \left(q^2+x_2\right)^k
  \nonumber\\
  &=&
  \left(q^2\right)^{\alpha_2+\alpha_3+\beta+k+2}
  \frac{\Gamma(\alpha_2+1)\Gamma(\alpha_3+1)
        \Gamma(-\alpha_2-\alpha_3-\beta-k-2)}
  {\Gamma(-\beta)}
  \nonumber\\&&\mbox{}
  \times \frac{\Gamma(-\alpha_3-\beta-1)}{\Gamma(-\alpha_3-\beta-k-1)}
  \,,
  \label{eq:Ih5}
\end{eqnarray}
turns out to be very useful.
In principle, the complete construction of $c_{71}$ requires the
proper linear combination with $c_{51}$. However, since we have
$c_{71}(\un_{51})=0$ this is not necessary.
For $c_{81}$ no separate calculation is necessary as the symmetry of
the basic polynomial can be exploited and one has
$c_{81}(n_1,n_2,n_3,n_4,n_5,n_6,n_7)=c_{71}(n_4,n_3,n_2,n_1,n_5,n_7,n_6)$.

The most complicated coefficient function is certainly $c_{41}$ since
there are four non-trivial integrations over $x_2$, $x_3$, $x_6$ and $x_7$
left. If $n_6$ or $n_7$ are less than or equal to zero the integrations can
be
performed in terms of
$\Gamma$ functions with the help of the formulae provided above.
However, for $n_6\ge1$ and $n_7\ge1$ this is not possible. In this
case the idea is to use IBP in order to reduce the four-parametric
integral representation
\begin{eqnarray}
  I_{41}^{A, \rm aux}(n_2,n_3,n_6,n_7,n_d) &=&
  \int\ldots\int \left[P_{41}(x_2,x_3,x_6,x_7)\right]^{z-n_d} \,
  \frac{\dd x_2 \, \dd x_3 \, \dd x_6 \, \dd x_7}
  { x_2^{n_2} x_3^{n_3} x_6^{n_6} x_7^{n_7} }
  \,,
  \label{auxint}
\end{eqnarray}
(with $z=(d-h-1)/2=d/2-5/2$)
to the auxiliary master integral $I_{41}^{A, \rm aux}(1,1,1,1,0)$.
$P_{41}$ is obtained from $P_A$ by setting $x_1$, $x_4$ and $x_5$ to zero.
We should stress that the corresponding recurrence procedure is
significantly simpler than the original one which involves seven
denominators. Furthermore, if during the recursion either $n_6$ or
$n_7$ becomes negative the corresponding expressions can immediately be
expressed in terms of $\Gamma$ functions.
The five IBP relations which are useful for the reduction to 
$I_{41}^{A, \rm aux}(1,1,1,1,0)$ can be obtained 
by either differentiating the
integrand with respect to $x_i$ ($i=2,3,6,7$) or by writing
$P_{41}^{z-n_d}=P_{41}^{z-n_d-1}P_{41}$ and inserting the explicit
result for the last factor.
The proper combination of these relations leads to new ones which
allows the following steps to be performed in an automatic way:
\begin{enumerate}
  \item
  Reduce $n_6$ and $n_7$ to one.
  \item
  Reduce $n_2,n_3>0$ to $n_2,n_3\le0$.
  \item
  Use IBP recurrence relations to obtain $n_2=n_3$.
  \item
  Reduce $n_2=n_3<0$ to $n_2=n_3=0$.
  \item
  Adjust the dimension, {\it i.e.}, reduce $n_d$ to zero.
\end{enumerate}
A simple relation transforms $I_{41}^{A, \rm aux}(0,0,1,1,0)$
to $I_{41}^{A, \rm aux}(1,1,1,1,0)$.
At this point one constructs the final coefficient function $c_{41}$
by considering the linear combination with $c_{51}$, $c_{71}$ and
$c_{81}$. Since one has $c_{41}(\un_{71})=c_{41}(\un_{81})=0$
we are left with
\begin{eqnarray}
  c_{41}(\un) &=& c^0_{41}(\un) - c^0_{41}(\un_{51}) c_{51}(\un)
  \,,
\end{eqnarray}
where
\begin{eqnarray}
  c^0_{41}(\un_{51}) &=&
  -\frac{1}{q^2 v^2}
  \frac{4 (d-3) (3d-14) (3d-10 ) (3d-8 )}
  {(d-4 )^2  (3d-13 ) (3d-11 )}
  \nonumber\\&&\mbox{}
  + \frac{(d-5)^2}{(3d-13 ) (3d-11 )} (q^2)^2 I_{41}^{A, \rm aux}(1,1,1,1,0)
  \,.
\end{eqnarray}
In this combination the auxiliary master integral
$I_{41}^{A, \rm aux}(1,1,1,1,0)$ cancels and $c_{41}(\un)$ is a rational
function in $d$.

The master integral $I_{61}$ forms a family by its own.
However, as the polynomial $P_{61}$ is quadratic in $x_7$
and thus the corresponding
recurrence relation shifts $n_7$ only in steps of two
it is necessary to introduce in addition the master integral
$I_{61a}$ where $n_7=-1$.
The practical implementation within our method does not need this recurrence
relation, however, feels the necessity of introducing $I_{61a}$.
The very calculation of the coefficient function is identical for $I_{61}$
and
$I_{61a}$. For $n_3\leq 0$ it can be done in terms of $\Gamma$ functions
with the
integration order $x_3$, $x_1$, $x_7$. On the other hand, for $n_3>0$ a
simple
one-step relation reduces $n_3$ to zero.

%%%%%%%%%%

\begin{figure}[t]
  \begin{center}
      \leavevmode
      \epsfxsize=16cm
      \epsffile[80 250 500 600]{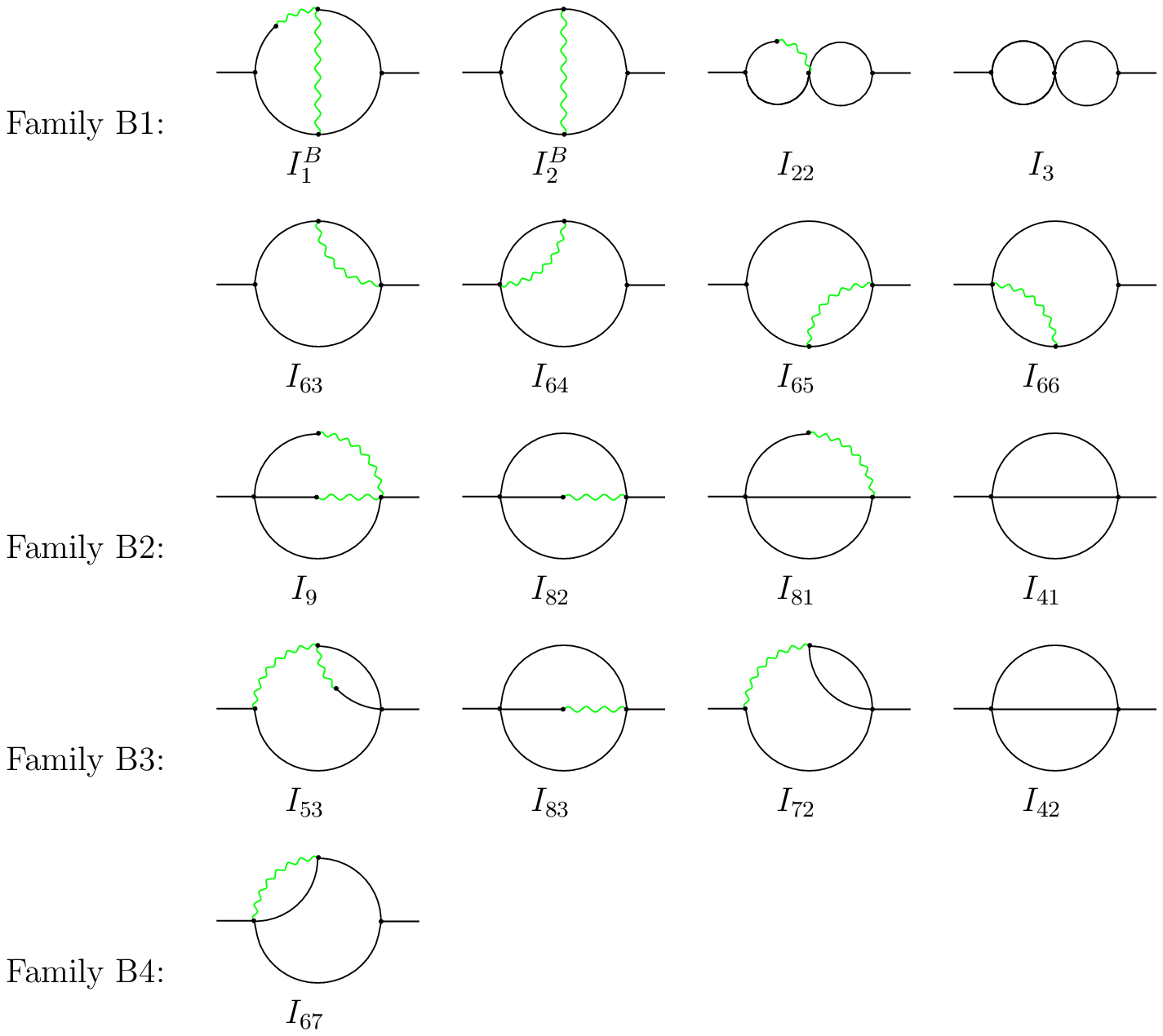}
  \end{center}
  \caption{\label{fig:potb}
    Feynman diagrams corresponding to the master integrals
    of case~B. In addition to $I_{6i}$ ($i=3,\ldots,7$)
    there are also master integrals
    ($I_{6ia}$) containing an irreducible numerators (see text).
          }
\end{figure}

Let us now come to case~B.
As one can see from Eq.~(\ref{eq:polb}) the basic polynomial is quite
similar to the one of case~A which can be used while computing the
coefficient functions. However, the symmetry
can only be exploited if $n_7\le0$ as for
$n_7>0$ the factor $(x_6-x_7)$ would appear in the denominator.

Altogether there are four families which, however, show a more
complicated structure than in case~A. More precisely one has
\begin{itemize}
\item Family~B1.
  There are twelve master integrals which obey the hierarchies
  $I^B_1 > \{I^B_{2},I_{22} \} > I_3$
  and $I^B_1 > I^B_{2} > \{I_{6i},I_{6ia}\}$ ($i=3,4,5,6$) and are given by
  \begin{eqnarray}
    I^B_1 &=&    F_B(1,1,1,1,0,1,1) \,,\nonumber\\
    I^B_{2} &=& F_B(1,1,1,1,0,0,1) \,,\nonumber\\
    I_{22} &=& F_B(1,1,1,1,0,1,0) \,,\nonumber\\
    I_3 &=& F_B(1,1,1,1,0,0,0) \,,\nonumber\\
    I_{63} &=& F_B(1,1,1,0,0,0,1) \,,\nonumber\\
    I_{64} &=& F_B(1,1,0,1,0,0,1) \,,\nonumber\\
    I_{65} &=& F_B(1,0,1,1,0,0,1) \,,\nonumber\\
    I_{66} &=& F_B(0,1,1,1,0,0,1)\,.
  \end{eqnarray}
  Furthermore there are master integrals with $n_6=-1$
  \begin{eqnarray}
    I_{63a} &=& F_B(1,1,1,0,0,-1,1)\,,\nonumber\\
    I_{64a} &=& F_B(1,1,0,1,0,-1,1)\,,\nonumber\\
    I_{65a} &=& F_B(1,0,1,1,0,-1,1)\,,\nonumber\\
    I_{66a} &=& F_B(0,1,1,1,0,-1,1)\,.
  \end{eqnarray}

\item Family~B2.
  One has four master integrals which obey the following hierarchy:
  $I_9 > \{I_{82},I_{81}\} > I_{41}$ with
  \begin{eqnarray}
    I_9 &=& F_B(1,0,0,1,1,1,1)\,,\nonumber\\
    I_{82} &=& F_B(1,0,0,1,1,0,1)\,,\nonumber\\
    I_{81} &=& F_B(1,0,0,1,1,1,0)\,,\nonumber\\
    I_{41} &=& F_B(1,0,0,1,1,0,0)\,.
  \end{eqnarray}

\item Family~B3.
  Similarly to Family~B2
  one has four master integrals obeying the hierarchy
  $I_{53} > \{I_{83},I_{72}\} > I_{42}$ with
  \begin{eqnarray}
    I_{53} &=& F_B(0,1,1,0,1,1,1)\,,\nonumber\\
    I_{83} &=& F_B(0,1,1,0,1,0,1)\,,\nonumber\\
    I_{72}  &=& F_B(0,1,1,0,1,1,0)\,,\nonumber\\
    I_{42} &=& F_B(0,1,1,0,1,0,0)\,.
  \end{eqnarray}

\item Family~B4 consists of the two master integrals
  \begin{eqnarray}
   I_{67} &=& F_B(0,1,0,1,1,1,0) \,,\nonumber\\
   I_{67a} &=& F_B(0,1,0,1,1,1,-1) \,,
  \end{eqnarray}
  which is similar to the Families~A4 and~A5 of case~A.

\end{itemize}
The construction of the coefficient functions $c_{1}^B$,
$c_{2}^B$ and $c_{22}$ from the family~B1
proceeds along the same lines as in case~A.
In the case of $c_3$ there is a slight complication as, in contrast to
case~A, $c_3(\un_1)\not=0$. As a consequence an auxiliary master
integral\footnote{The definition of $I_{3}^{B, \rm aux}(n_5,n_6,n_7,n_d)$
  is in analogy to Eq.~(\ref{auxint}).},
$I_{3}^{B, \rm aux}(0,1,1,0)$
has to be introduced which is only canceled after considering
the proper linear combination with $c_1$.
The reduction to $I_{3}^{B, \rm aux}(0,1,1,0)$ is straightforward.

Family~B1 has four more members, $I_{63}$, $I_{64}$, $I_{65}$ and
$I_{66}$, which belong to the four hierarchies $I_1^B>I_{2}^B>I_{6i}$
($i=3,4,5,6$). Thus, in order to obtain the coefficient functions
$c_{6i}$ one has to consider the linear combination
\begin{eqnarray}
  c_{6i} &=& c^0_{6i}
           - c^0_{6i}(\un_1^B) c_1^B(\un)
           - c^0_{6i}(\un_2^B) c_2^B(\un)
  \,.
  \label{eq:c6i}
\end{eqnarray}
Let us in the following restrict the discussion to $c_{63}$ since the
results for the other three coefficients can be obtained by exploiting
the symmetry.
$c^0_{63}$ is given by an integral representation of the
form
\begin{eqnarray}
  c^0_{63} &\sim&
  \int\ldots\int \left[P_{63}(x_4,x_5,x_6)\right]^{z-n_d} \,
  \frac{\dd x_4 \, \dd x_5 \, \dd x_6}
  { x_4^{n_4} x_5^{n_5} x_6^{n_6} }
  \label{eq:intrep63}
  \,,
\end{eqnarray}
with
\begin{eqnarray}
  P_{63} &=&
  (q^2)^2 v^2 x_5
  +q^2 v^2 \left(x_4 x_5 - x_5^2\right)
  -4q^2x_5 x_6^2
  -x_4^2 x_6^2
  \label{eq:P63}
  \,.
\end{eqnarray}
For $n_4\le0$, where we have $c_1^B(\un)=c_2^B(\un)=0$,
the integrals in Eq.~(\ref{eq:intrep63}) can be solved
analytically in the order $x_4$, $x_5$, $x_6$ using
Eq.~(\ref{eq:Ih4}) for $x_4$, the formula
\begin{eqnarray}
  I_h^{(6)}(\alpha,\beta)
  &=& \int_0^\infty  {\rm d}x\, x^\alpha \left(x+a\right)^\beta
  \nonumber\\
  &=& a^{\alpha+\beta+1}
  \frac{\Gamma(1+\alpha)\Gamma(-\alpha-\beta-1)}{\Gamma(-\beta)}
  \,,
  \label{eq:Ih6}
\end{eqnarray}
for $x_5$ and Eq.~(\ref{eq:Ih3}) extended to non-integer $k_3$ for $x_6$.
For $n_4>0$ two auxiliary master
integrals,
$I_{63}^{B, \rm aux}(1,0,0,0)$ and $I_{63}^{B, \rm aux}(1,0,1,0)$,
have to be introduced where the reduction of
Eq.~(\ref{eq:intrep63}) proceeds s follows
\begin{enumerate}
  \item
  Reduce $n_4$ to one.
  \item
  Reduce $n_5$ to zero.
  \item
  The reduction of $n_6$ can only be performed in steps of two. Thus
  one end up with $n_6=0$ or $n_6=-1$.
  \item
  Adjust the dimension, {\it i.e.}, reduce $n_d$ to zero.
\end{enumerate}
The corresponding recurrence relations are
easily derived from Eq.~(\ref{eq:P63}).
It is interesting to note that in Eq.~(\ref{eq:c6i})
the master integral $I_{63}^{B, \rm aux}(1,0,1,0)$ is canceled from
$c_1^B$ and $I_{63}^{B, \rm aux}(1,0,0,0)$ from $c_2^B$.
Please note that due to the structure of Eq.~(\ref{eq:P63}) in addition to
$I_{63}$ also a master integral with $n_6=-1$, $I_{63a}$, has to be
introduced which, however, has the same coefficient function as $I_{63}$.
We want to mention that for $c_{63}$ and $c_{65}$ the master integrals
$I_{6}$ and $I_{6a}$ are needed while for $c_{64}$ and $c_{66}$ the
integrals
$I_{6}$ and $I^B_{6a}$ are necessary.

As far as the families~B2,~B3 and~B4 are concerned the strategy
to construct the coefficient function goes along the same lines as for
the families~A2,~A3 and~A4.

For completeness let us in the following list all occurring
master integrals. They have been obtained with the help
of the program package developed for the
calculation performed in Ref.~\cite{KPSS1} where conventional recurrence
relations have been applied.
In the results the overall factor $\left(i \pi^{d/2}\right)^2 Q^{-4\ep}$
is omitted and $Q=\sqrt{-q^2}$ has been introduced.
Moreover the standard factor $\e^{-2\gm_{\rm E}\ep}$ is
pulled out in the formulae where an expansion in $\ep$ is performed.
We obtain
\begin{eqnarray}
  I_1 &=& \frac{\pi}{Q^2 v^2}
  \frac{\Gm^2(5/2 - d/2) \Gm^4(d/2-3/2)}
  {\Gm^2(d-3)}
  \,, \nn \\ %  \hspace*{-50mm}
  I_2 &=& - \frac{\sqrt{\pi}}{Q v}
  \frac{ \Gm(2 - d/2) \Gm(5/2 - d/2)\Gm^2(d/2-1) \Gm^2(d/2-3/2)}
  {\Gm(d-3 ) \Gm(d-2)}
  \,, \nn \\ %&&  \hspace*{-50mm}
  I_3&=& \frac{\Gm^2(2 - d/2) \Gm^4(d/2-1)}{\Gm^2(d-2)}
  \,, \nn \\ %&&  \hspace*{-50mm}
  I_4 &=& - Q^2
  \frac{\Gm(3 - d) \Gm^3(d/2-1)}{\Gm(3 d/2-3)}
  \,, \nn \\ %&&  \hspace*{-50mm}
  I_5 &=&
  \frac{\pi^2}{v^2}\left[ -\frac{2}{3\ep} - 4  +(-24 + \frac{7}{9} \pi^2)
\ep
    + {\cal O}(\ep^2)
  \right]
  \,, \nn \\ %&&  \hspace*{-50mm}
  I_6 &=& \frac{\sqrt{\pi} Q}{v}
  \frac{2^{d-2} \Gm(3 - d) \Gm(7/2 - d) \Gm(d/2-1) \Gm^2(d-5/2)}
  {\Gm(2 - d/2) \Gm(2 d-5)}
  \,, \nn \\ %&&  \hspace*{-50mm}
  I_{6a} &=& - \sqrt{\pi} Q^2
  \frac{2^{d-2}  \Gm^2(3 - d) \Gm(d/2-1) \Gm^2(d-2)}
  {\Gm(3/2 - d/2) \Gm(2 d-4)}
  \,, \nn \\ %&&  \hspace*{-50mm}
  I_7 &=&  \frac{\sqrt{\pi} Q}{v}
  \frac{ \Gm(7/2 - d) \Gm^2( d/2-1) \Gm(d/2-3/2) \Gm(d-5/2 )}
  {\Gm(d-2) \Gm(3 d/2-4)}
  \,, \nn \\ %&&  \hspace*{-50mm}
  I_8 &=& I_7
  \,, \nn \\ %&&  \hspace*{-50mm}
  I_9 &=& I_5
  \,, \nn \\ %&&  \hspace*{-50mm}
  I^B_1 &=&  \frac{1}{2}I_1
  \,, \nn \\ %&&  \hspace*{-50mm}
  I^B_2 &=&
  \frac{\pi ^2 }{Q v}\left[
    -4 \ln 2 + \ep \left(
      \frac{5}{3} \pi^2 - 16 \ln 2 - 4\ln^2 2
    \right)+ {\cal O}(\ep^2)\right]
  \nn \\ %&&  \hspace*{-50mm}
  I^B_{6a} &=& -I_{6a}
  \;.
\label{masters}
\end{eqnarray}
Observe that $I_5 = I_9$ and $I_7 = I_8$
although the corresponding integrands are
certainly different.
These equalities can be immediately be seen by a simple
change of variables.
Because of $I_7 = I_8$ we have in both cases
one master integral less and end up with
eight master integrals, $I_1,\ldots,I_7$ and $I_{6a}$, in case~A
while in case~B ten master integrals contribute,
$I_2, \ldots,I_7, I_9, I^B_{6a},I^B_1$ and $I^B_2$.
Please note that only two of them are not known
in terms of $\Gamma$ functions. In the above list their result is given as
an
expansion in $\ep$ up to terms of order $\ep^1$.

A realistic diagram (in particular for arbitrary gauge parameter $\xi$)
does not immediately lead to the integrals $F_A$ and $F_B$ defined in
Eqs.~(\ref{2lpA}) and~(\ref{2lpB}). However, it is straightforward to map
the
expressions to the function (see also the appendix of
Ref.~\cite{Kniehl:2002br})
\begin{eqnarray}
  J(\un) &=&
  \int\int\frac{{\rm d}^dk{\rm d}^dl}{(k^2)^{n_1}(l^2)^{n_2}
    [(k-q)^2]^{n_3}[(l-q)^2]^{n_4}[(k-l)^2]^{n_5}(v \cdo k)^{n_6}
    (v \cdo l)^{n_7}  [v \cdo (k-l)]^{n_8} }
  \,,
  \nonumber\\
\end{eqnarray}
with the obvious identities
\[J(n_1,n_2,n_3,n_4,n_5,n_6,n_7,0)=F_A(\un),\;\;
J(n_1,n_2,n_3,n_4,n_5,n_6,0,n_8)=F_B(\un).\]
The reduction to $F_A$ and $F_B$ is achieved with the help
of the relation
\begin{eqnarray}
  J(\un) &=&
  J(n_1,n_2,n_3,n_4,n_5,n_6+1,n_7-1,n_8)
  + J(n_1,n_2,n_3,n_4,n_5,n_6+1,n_7,n_8-1)
  \,,
  \nonumber\\
\end{eqnarray}
where either $n_7$ or $n_8$ are reduced to zero from positive
values. In the case of negative $n_7$ or $n_8$ such a reduction
is similar.

The method described above has been implemented in a {\tt Mathematica}
package. The typical runtime of an integral contributing to the static
potential or to the $1/(m_q r^2)$ corrections
amounts to a few seconds.
This extends to the order of a minute for integrals which occur in the
calculation for a general gauge parameter where the exponents of the
propagators can be significantly higher.

Let us at the end of this section present some explicit results
which illustrate the functionality of our method.
For this reason we list next to the results for
$F_A(\un)$ and $F_B(\un)$ also the (non-zero) expressions for
the individual coefficient functions.

For $\un=(2,2,1,1,1,1,1)$ we get\footnote{The same standard overall
factors are pulled out as in Eqs.~(\ref{masters}).}
\begin{eqnarray}
  F_A(2,2,1,1,1,1,1) &=& \frac{1}{(q^2)^4 v^2}
  \left(
    \frac{2}{3\ep} + \frac{4}{3\ep}\pi^2
    -\frac{16}{9}
    + \frac{368}{45}\pi^2
    -8\zeta(3)
    +{\cal O}(\ep)
  \right)
  \,,
\end{eqnarray}
with the coefficients
\begin{eqnarray}
&&c_1= \frac{2 (d - 5) (d - 4)}
{q^6}
\,,\quad
c_3= \frac{8 (d - 5) (d - 3)^2}
{(d - 4) q^8 v^2}
\,,\quad
\nonumber\\&&
c_{41}= c_{42}=
%\nonumber\\&&
\frac{-3(d - 3)(3d - 16)(3d - 14)(3d - 10)(3d - 8)
  ( 5d^3- 93d^2 + 588d-1264  )}
{(d - 9)(d - 8)(d - 7)(d - 6)^2
  (d - 4)^2q^{10}v^2}
\,,\nonumber\\&&
c_{51}= c_{52}=
\frac{-3 (3d-17) (3d-13) (3d-11)}
{(d - 9) (d - 7) q^8}
\,,\nonumber\\&&
c_{61}= c_{62}=
\frac{-32 (2 d -13 ) (2 d-11) (2 d-9) (2 d-7) (2 d-5)}
{(d - 9) (d - 7) (d - 6) (d - 4) q^{10} v^2}
\,,
\end{eqnarray}
where $c_{61}$ and $c_{62}$ get multiplied by $I_{6a}$.
In the case~B the results read
\begin{eqnarray}
  F_B(2,2,1,1,1,1,1) &=& \frac{1}{(q^2)^4 v^2}
  \left(
  -\frac{1}{3\ep} + \frac{4}{3\ep}\pi^2
+  \frac{8}{9}
  + \frac{368}{45}\pi^2
  + 4\zeta(3)
  +{\cal O}(\ep)
  \right)
  \,,
\end{eqnarray}
\begin{eqnarray}
&&
c_1^{B}= \frac{2 (d - 5) (d - 4)}
{q^6}
\,,\quad
c_3=
\frac{-4 (d - 5) (d - 3)^2}{(d - 4) q^8 v^2}
\,,\quad
\nonumber\\&&
c_{41}=
%\nonumber\\&&
\frac{3 (d - 3) (3 d -16) (3 d -14) (3 d -10) (3 d -8)
  (7 d^3- 117 d^2+ 654 d -1232  )}
{(d - 9) (d - 8) (d - 7) (d - 6)^2
  (d - 4)^2 q^{10} v^2}
\,,\nonumber\\&&
c_{42}=
\frac{-6 (d - 3) (3 d -16) (3 d -14) (3 d -10) (3 d -8)
  ( d^3- 12 d^2+ 33 d +16  )}
{(d - 9) (d - 8) (d - 7) (d - 6)^2
  (d - 4)^2 q^{10} v^2}
\,,\nonumber\\&&
c_{53}= \frac{-3 (3 d -17) (3 d -13) (3 d -11)}
{(d - 9) (d - 7) q^8}
\,,\nonumber\\&&
c_{63}= c_{64}=
- \frac{4 (2 d -7) (2 d -5)
(15 d^4- 304 d^3 + 2240 d^2- 7093 d  +8118 )}
{(d - 9) (d - 7) (d - 6) (d - 4) q^{10} v^2}
\,,\nonumber\\&&
c_{65}= c_{66}=
\frac{4 (2 d -7 ) (2 d -5) (d^2- 17 d +55 )}
{(d - 7) (d - 4) q^{10} v^2}
\,,\nonumber\\&&
c_{67}= \frac{-32 (2 d -13) (2 d -11) (2 d -9) (2 d -7) (2 d -5)}
{(d - 9) (d - 7) (d - 6) (d - 4) q^{10} v^2}
\,,\nonumber\\&&
c_9= \frac{-3 (3 d -17) (3 d -13) (3 d -11)}
{(d - 9) (d - 7) q^8}
\,,
\end{eqnarray}
where $c_{63}$ and $c_{65}$ get multiplied by $I_{6a}$ and
$c_{64}$, $c_{66}$ and $c_{67}$ by $I_{6a}^B$.
As a second example let us consider $\un=(1,1,2,1,1,-1,1)$
which gives
\begin{eqnarray}
  F_A(1,1,2,1,1,-1,1) &=& \frac{1}{(q^2)^2}
  \left(
    -\frac{1}{2\ep}
    +\frac{3}{2}
    -2\zeta(3)
    +{\cal O}(\ep)
  \right)
  \,,
\end{eqnarray}
\begin{eqnarray}
&&
c_3= \frac{2 (d - 3)}
{(d - 4) q^4}
\,,\quad
c_{41}= \frac{-3 (3 d -10) (3 d -8) (d^2- 5 d +2 )}
{2 (d - 6) (d - 5) (d - 4)^2 q^6}
\,,\nonumber\\&&
c_{42}= \frac{3 (d - 5) (d-2) (3 d -10) (3 d -8)}
{2 (d - 6) (d - 4)^2 q^6}
\,,\nonumber\\&&
c_{62}= \frac{4 (2 d -9) (2 d -7) (2 d -5)}
{(d - 5) (d - 4) q^6}
\,,
\end{eqnarray}
where $c_{62}$ gets multiplied by $I_{6a}$.
\begin{eqnarray}
  F_B(1,1,2,1,1,-1,1) &=& \frac{1}{(q^2)^2}
  \left(
    -\frac{1}{2\ep}
    +\frac{1}{2}
    +{\cal O}(\ep)
  \right)
  \,,
\end{eqnarray}
\begin{eqnarray}
&&
c_3= \frac{(d - 5) (d - 3)}
{(d - 6) q^4}
\,,\quad
c_{41}= \frac{-3 (3 d -10) (3 d -8) ( d^2- 9 d +22 )}
{2 (d - 6)^2 (d - 5) (d - 4) q^6}
\,,\nonumber\\&&
c_{42}= \frac{3 (3 d -10) (3 d -8) (d^2- 11 d +26 )}
{2 (d - 6)^2 (d - 4) q^6}
\,,\quad
c_{63}= \frac{(2 d-11) (2 d-7) (2 d-5)}
{(d - 6) (d - 5) q^6}
\,,\nonumber\\&&
c_{64}= \frac{-(2 d-7) (2 d-5)}
{(d - 6) (d - 5) q^6}
\,,\quad
c_{65}= \frac{(2 d-7)^2 (2 d-5)}
{(d - 6) (d - 5) q^6}
\,,\nonumber\\&&
c_{66}= \frac{-(2 d-7) (2 d-5) (4 d-19)}
{(d - 6) (d - 5) q^6}
\,,
\end{eqnarray}
where again
$c_{63}$ and $c_{65}$ get multiplied by $I_{6a}$ and
$c_{64}$ and $c_{66}$ by $I_{6a}^B$.

%%%%%%%%%%%%%%%%%%%%%%%%%%%%%%%%%%%%%%%%%%%%%%%%%%%%%%%%%%%%

\section{\label{sec:concl}Conclusion}

In this paper we described an algorithmic procedure to identify the master
integrals for a given class of Feynman diagram.
Furthermore details are given for the computation of the
corresponding coefficient functions which are formally defined in
Eq.~(\ref{eqsol}).
The general procedure was applied to useful one- and two-loop integrals
which often appear in practical applications. In particular we discussed in
great detail the integrals needed for the calculation of the two-loop heavy
quark potential.
We do hope that, with our instructions, the reader will be able to
apply Eq.~(\ref{eqsol}) to various recursion problems for Feynman
integrals.

Let us finally point out that there is also
another approach to use Eq.~(\ref{eqsol}) which
is based on an expansion of some parametric representations
(originating from Eq.~(\ref{eqsol})) in
the limit of large dimension $d$~\cite{Bai8}. There the crucial assumption
is the rational dependence of the coefficients $c_i$ on $d$. No
description of this procedure has been published up to now.

%%%%%%%%%%%%%%%%%%%%%%%%%%%%%%%%%%%%%%%%%%%%%%%%%%%%%%%%%%%%

\section*{Acknowledgements}

M.S. would like to thank Y.~Schr\"oder for discussions.
The work of V.S. was supported by the Russian Foundation for Basic
Research through project 01-02-16171,
the Volkswagen Foundation through contract No.~I/77788, and
INTAS through grant 00-00313.

\end{document}